\documentclass[conference,letterpaper]{IEEEtran}
\IEEEoverridecommandlockouts                              

\usepackage{fancyhdr}

\usepackage{graphicx}
\usepackage{caption}
\usepackage{cite}
\usepackage{commath}
\usepackage{subcaption}

\graphicspath{ {Figures/} }

\ifCLASSINFOpdf
\else
\fi
\title{\LARGE \bf
Adaptive Twisting Sliding Mode Control for \\Quadrotor Unmanned Aerial Vehicles
}
\author{\IEEEauthorblockN{V.T. Hoang, M.D. Phung,  Q.P. Ha }
\thanks{This work was supported by a UTS FEIT 2017 Data Arena research grant.}
\thanks{The authors are with School of Electrical and Data Engineering, Faculty of Engineering and Information Technology (FEIT), University of Technology Sydney (UTS), 81 Broadway, Ultimo NSW 2007, Australia
        {\tt\small \{VanTruong.Hoang, ManhDuong.Phung, Quang.Ha\}@uts.edu.au}}%
}

\begin{document}
\bibliographystyle{IEEEtran}



\maketitle

\thispagestyle{fancy}
\fancyhead{}
\lhead{}
\lfoot{}
\cfoot{}
\rfoot{}
\renewcommand{\headrulewidth}{0pt}
\renewcommand{\footrulewidth}{0pt}

\begin{abstract}
This work addresses the problem of robust attitude control of quadcopters. First, the mathematical model of the quadcopter is derived considering factors such as nonlinearity, external disturbances, uncertain dynamics and strong coupling. An adaptive twisting sliding mode control algorithm is then developed with the objective of controlling the quadcopter to track desired attitudes under various conditions. For this, the twisting sliding mode control law is modified with a proposed gain adaptation scheme to improve the control transient and tracking performance. Extensive simulation studies and comparisons with experimental data have been carried out for a Solo quadcopter. The results show that the proposed control scheme can achieve strong robustness against disturbances while is adaptable to parametric variations.


\end{abstract}

\IEEEpeerreviewmaketitle

\section{Introduction}
Over the last decade, the quadcopter unmanned aerial vehicle (UAV) has received much research attention. UAVs or drones nowadays have found various applications, ranging from military to industry for surveillance and rescue, civil infrastructure monitoring and inspection. The development of UAVs also covers many areas, including mechatronics and robotics research, control and planning, data engineering and communication, see. e.g., \cite{Derafa:2012}, \cite{Chen:2015}, \cite{Phung:2016}. While UAV applications continue to grow, a great deal of effort is being devoted to better handle the control problem of quadcopters to cope with the complexity of their dynamics, system parameter variations and particularly, large external disturbances. A quadrotor drone has generally six degrees of freedom but only four independent inputs, i.e., the four rotor speeds, thus making it an underactuated system. Apart from the coupling condition of rotational and translational motion, UAVs are also subject to highly nonlinear dynamics and aerodynamic effects, which cause microscopic frictions acting on the quadcopter, leading to the need to generate compensative forces to maintain proper movements at the steady state. Designing robust control algorithms for quadcopters is therefore an interesting topic.

In the literature, several control algorithms have been developed for quadcopters such as command-filtered PD/PID control \cite{Zuo:2010}, integral predictive/$H_\infty$ control \cite{Raffo:2010}, optimal control \cite{Ritz_2011}, and extended potential field \cite{La_2016}. Among the robust control techniques developed for UAV, the sliding mode control (SMC) is widely used due to its salient capability of maintaining system performance against the influence of modelling errors and external disturbances \cite{Xu:2006, Besnard:2007, Derafa:2012}. In SMC, the chattering effect occurring in the steady state usually excites unmodeled frequencies of the system dynamics. Higher-order sliding modes (HOSM)  based on a higher-order derivative of the sliding function have been introduced to reduce this effect \cite{Levant:1993, Manceur:2012, Rubio:2014} and also to improve the finite-time convergence \cite{Utkin:2016}. 

In the HOSM control framework, most popular are twisting controllers \cite{ Polyakov:2009} and their modified versions like super-twisting \cite{ Polyakov:2009R, Ha:2013}, adaptive twisting \cite{Shtessel:2012}, and accelerated twisting \cite{Dvir:2015}. Owing to their advantages, these HOSM techniques have been applied to UAV control \cite{Zheng:2014, Hoang:2017}. However, these control laws are indeed complicated and would require some simplification. To this end, the one-stage algorithm of the accelerated twisting sliding mode (ATSM), where the control gain is modified to be always greater than an exponential function of the sliding function magnitude, appears not too complicated but can guarantee accelerated finite-time, or at least, fixed-time convergence \cite{Dvir:2015}. Motivated by the work therein, we propose in this paper an adaptive scheme to be able to adjust the control gain of the twisting control law and apply it to control the attitude of quadcopters in harsh conditions with nonlinearity, external disturbances, uncertain dynamics and strong coupling. The control performance of the proposed controller is verified in simulation and also by comparison with real-time data of a Solo drone.

The paper is organised as follows. Section \ref{dynamics} briefly describes the dynamic model of the quadcopter. Section \ref{controller} presents the design of the proposed adaptive twisting sliding mode controller. Simulation and comparison with experimental data are introduced in Section \ref{results}. The paper ends with a conclusion and recommendation for future work.

\section{Dynamic model} \label{dynamics}
\begin{figure}[!htb]
    \centering 
    \includegraphics[width=9cm]{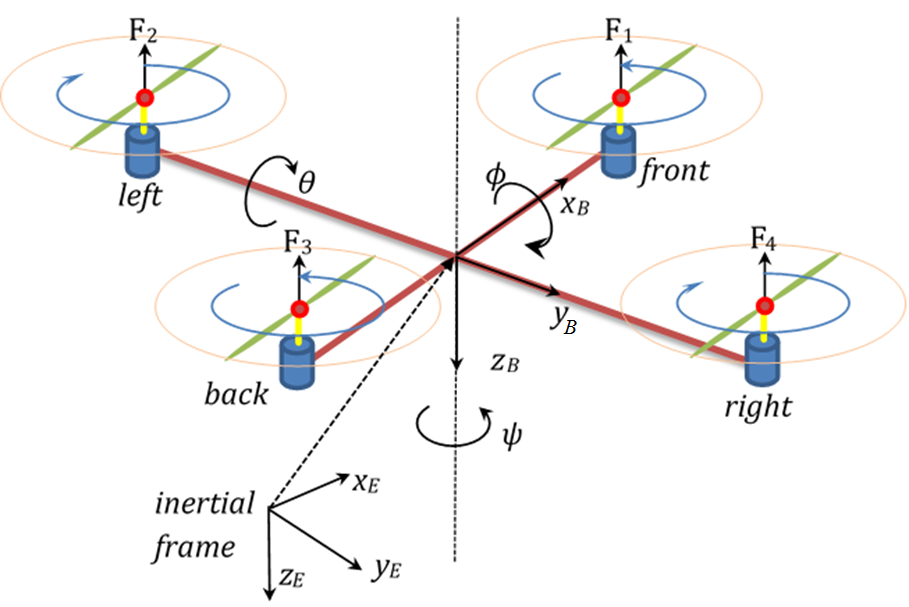}
    \caption{A schematic diagram of quadcopter.}\label{Fig1}
\end{figure}

The model of the quadcopter used in this work is illustrated in Fig. \ref{Fig1}, wherein the inertial frame, $(x_E, y_E, z_E)$, is defined by the ground with the $z$ axis being directed down to the earth centre, and the body frame, $(x_B, y_B, z_B)$, is specified by the orientation of the quadcopter with the $z$ axis also pointing downward and the $x$ and $y$ axes pointing to the arms' directions.

The translational motion of the quadcopter in the inertial frame is determined by its position, $\xi=({x , y , z})^T$, and velocity, $\dot{\xi}=({\dot x , \dot y , \dot z})^T$. The UAV attitude is described by Euler angles roll, pitch, and yaw,  $\Theta = ({\phi , \theta , \psi})^T$ with the corresponding angular rates $\dot \Theta =  ({\dot \phi , \dot \theta , \dot \psi})^T$. Let $\omega = [p,q,r]^T$ be the angular rate of the quadcopter in the inertial frame, i.e.:
\begin{equation}\label{Eq1}
\omega = \left [\begin{array}{ccc} 
    1 &0 & -s_\theta \\
    0 & c_\phi & c_\theta s_\phi \\
    0 & -s_\phi & c_\theta c_\phi 
\end{array} \right ]\dot\Theta ,
\end{equation}
where $s_x$ denotes sin$(x)$ and $c_x$ denote cos$(x)$. The transformation from the body to earth frames is then determined by the following rotation matrix:
\begin{equation}\label{Eq2}
R = \left [\begin{array}{ccc} 
c_\psi c_\theta & c_\psi s_\theta s_\phi-s_\psi c_\phi & c_\psi s_\theta c_\phi+s_\psi s_\phi\\
s_\psi c_\theta & s_\psi s_\theta s_\phi+c_\psi c_\phi & s_\psi s_\theta c_\phi-c_\psi s_\phi\\
-s_\theta & c_\theta s_\phi & c_\theta c_\phi
\end{array} \right].
\end{equation}

Since only the attitude control is concerned in this work, torque components for orientation of the UAV are considered here. They include the torque caused by thrust forces $\tau$, by body gyroscopic effects $\tau_b$, by propeller gyroscopic effects $\tau_p$, and by aerodynamic friction $\tau_a$. Components of the torque vector  $\tau = [\tau_\phi \enskip \tau_\theta \enskip  \tau_\psi]^T$, corresponding to rotation in the roll, pitch and yaw directions, are determined by:
\begin{align}\label{eq:torque_roll}
\tau_\phi &= l(F_2 - F_4),\\\label{eq:torque_pitch}
\tau_\theta &= l(-F_1 + F_3),\\\label{eq:torque_yaw}
\tau_\psi &= c(-F_1+F_2 -  F_3+F_4),
\end{align}
where $l$ is the distance from the motor to the centre of mass of the quadcopter, and $c$ is the force-to-torque coefficient. The body gyroscopic torque is determined by:
\begin{equation}
\tau_b = -S(\omega) I\omega,
\end{equation}
where $S(\omega)$ is a skew-symmetric matrix
\begin{align}\label{Eq5a}
S(\omega) = \left[ \begin{array}{ccc}
0 & -r & q\\ r & 0 & -p\\ -q & p & 0
\end{array}\right].
\end{align}
For small rotation angles of the quadrotor, $\omega$ is approximate to $\dot{\Theta}$. The attitude dynamic model of the quadcopter thus can be described as:
\begin{equation}\label{Eq5}
I\ddot{\Theta} = \tau_b  + \tau + \tau_p - \tau_a,
\end{equation}
where $I = \text{diag}[I_{xx},  I_{yy}, I_{zz}]$ is the matrix of inertia of the quadrotor, assumed to be symmetrical. 

In our system, the gyroscopic and aerodynamic torques are considered as external disturbances. 
Thus, the control inputs mainly depend on the thrust torque  $\tau= [\tau_\phi \enskip \tau_\theta \enskip  \tau_\psi]^T$. From (\ref{eq:torque_roll}), (\ref{eq:torque_pitch}) and (\ref{eq:torque_yaw}), the control inputs can be described as:
\begin{align}\label{Eq6}
\left[\begin{array}{c} u_\phi\\ u_\theta\\ u_\psi \\ u_z\end{array}\right] = \left[\begin{array}{c} \tau_\phi\\ \tau_\theta\\  \tau_\psi \\ F\end{array}\right]  = \left[\begin{array}{cccc}
0 & l & 0 & -l \\	-l & 0 & l & 0\\ -c & c & -c & c \\ 1 & 1 & 1 & 1
\end{array}\right] \left[\begin{array}{c} F_1 \\ F_2\\ F_3\\ F_4\end{array}\right],
\end{align}
where $F$ is the UAV lift, $u_z$ represents the total thrust acting on the four propellers and $u_\phi$, $u_\theta$ and  $u_\psi$ respectively represent the roll, pitch and yaw torques. As only the attitude of the quadcopter will be controlled, $u_z$ is assumed to balance with  the gravity. Therefore, the second-order nonlinear dynamic equations of the quadcopter for attitude control can be described by:
\begin{align}  \label{Eq6d}
\ddot{\phi} &= \dfrac{1}{I_{xx}}\left[(I_{yy} - I_{zz}) qr + u_\phi + d_\phi \right]\\  \label{Eq6e}
\ddot{\theta} &= \dfrac{1}{I_{yy}}\left[(I_{zz} - I_{xx}) pr + u_\theta + d_\theta\right]\\ \label{Eq6f}
\ddot{\psi} &= \dfrac{1}{I_{zz}}\left[(I_{xx} - I_{yy}) pq + u_\psi + d_\psi\right],   
\end{align}
where $ d_\phi, d_\theta$ and $d_\psi$ are angular acceleration disturbances. The quadcopter dynamics can be then represented as follows:
\begin{align}\label{Eq7b}
\begin{cases}
        	\dot{X}_1 = X_2 \\
        	\dot{X}_2 = I^{-1}\left[f(X) + u + d\right],
    	\end{cases}
\end{align}
where $X_1 = \Theta$, $X_2 = \dot{\Theta}$, $X=[X_1, X_2]^T$ is the state vector, $u = [u_\phi, u_\theta, u_\psi]^T$ is the input vector, $d = [d_\phi, d_\theta, d_\psi]^T$ is the disturbance vector, and $f(X)$ is the matrix represented as
\begin{align}\label{Eq8}
f(X) &= \left( \begin{array}{c}
(I_{yy} - I_{zz}) qr\\
(I_{zz} - I_{xx}) pr\\
(I_{xx} - I_{yy}) pq
\end{array} \right) .
\end{align}

In our system, the following assumptions are made:
\begin{itemize}
\item[A.1] The quadcopter structure is rigid and symmetric. 
\item[A.2] The reference trajectories and their first and second time derivatives are bounded.
\item[A.3] The velocity and the acceleration of the quadcopter are bounded.
\item[A.4] The orientation angles are limited to $\phi \in \left[-\dfrac{\pi}{2}, \dfrac{\pi}{2}\right]$, $\theta \in \left[-\dfrac{\pi}{2}, \dfrac{\pi}{2}\right]$ and $\psi \in \left[-\pi, \pi\right]$.
\end{itemize}
\section{Control Design} \label{controller}
Given the desired angle reference $X_{1d} = \{\phi_d, \theta_d, \psi_d\}^T$, the overall control law is proposed as:
\begin{equation}\label{Eq11}
u(t)= u_{eq}(t) + u_D(t),
\end{equation}
where $u_{eq}(t) = (u_{eq,i})^T$ and $u_D(t) = (u_{D,i})^T$, $i = 1, 2,3$, are respectively the equivalent control and the discontinuous part containing switching elements. In our system, the sliding surface equation is chosen as:
\begin{equation}\label{Eq10}
\mathbf{\sigma = \dot{e}} +\Lambda \mathbf{e},
\end{equation}
where $\Lambda = \text{diag}(\lambda_\phi, \lambda_\theta, \lambda_\psi)$ is a positive definite matrix being designed, and $\mathbf{e}$ is the control error, $\mathbf{e} = X_1 -X_{1d}$.

\subsubsection{Design $u_{eq}$}
The equation (\ref{Eq10}) can be rewritten for the attitude sliding surface as: 
\begin{equation}\label{Eq13}
\sigma = (\dot{X}_1 - \dot{X}_{1d})+\Lambda(X_1 - X_{1d}).
\end{equation}
Taking the time derivative of $\sigma$, we have:
\begin{equation}\label{Eq14a}
\dot{\sigma} = (\ddot{X}_1 - \ddot{X}_{1d})+ \Lambda(\dot{X}_1 - \dot{X}_{1d}),
\end{equation}
or
\begin{equation}\label{Eq14b}
\dot{\sigma} = -\ddot{X}_{1d} + \dot{X}_2+ \Lambda\dot{e}.
\end{equation}
Substituting $\ddot{X}$ from (\ref{Eq7b}) to (\ref{Eq14b}) yields: 
\begin{equation}\label{Eq15}
\dot{\sigma} = -\ddot{X}_{1d} + I^{-1}\left[ f(X) + u\right] + \Lambda\dot{e}.
\end{equation}
When the sliding mode has been induced, $u$ can be considered as the equivalent control $u_{eq}$. By driving the derivative of sliding surface to zero, the equivalent control rule can be obtained as follows:
\begin{equation}\label{Eq16a}
u_{eq} = I\left(\ddot{X}_{1d}  - \Lambda\dot{e}\right)- f(X).
\end{equation}
\subsubsection{Design $u_D$}
The discontinuous control is 
\begin{equation}\label{Eq16a1}
u_{D} = u_T,
\end{equation}
where the twisting controllers $u_{T,i}, ~i=1,2,3$ are adopted here as:
\begin{align}\label{Eq17}
u_{T,i} = \begin{cases} 
-\mu_i \alpha_i \text{sign}(\sigma_i) & \text{if} \enskip \sigma_i\dot{\sigma}_i \leq 0 \\
-\alpha_i \text{sign}(\sigma_i) & \text{if} \enskip \sigma_i\dot{\sigma}_i > 0,
\end{cases}
\end{align}
where $\mu_i<1$ is a fixed positive number and $\alpha_i >0$ is the control gain \cite{Levant:1993}. To improve the control transient and tracking performance, the gain $\alpha_i$ in (\ref{Eq17}) could be selected to satisfy the following condition for the one-stage accelerated twisting algorithm  \cite{Dvir:2015}:
\begin{align}
\alpha_i = \text{max} \{ \alpha_{*,i}, \gamma_i \abs{\sigma_i}^{\rho_i}\},
\end{align}
 where $\alpha_{*,i}$, $\gamma_i$ and $\rho_i$ are positive constants. 
Given that fixed time stability is required over a large operational region of the UAV, and motivated by the simplicity of the one-stage accelerated twisting algorithm mentioned above, we propose to adjust the gain $\alpha_i$ in (\ref{Eq17}) adaptively as in \cite{Plestan:2010,Hoang:2017}, to be constructed based on the following equation:
\begin{align}\label{Eq24}
\dot{\alpha}_i &= \begin{cases}
\bar{\omega_i} \abs{\sigma_i(\omega,t)} \text{sign}(|\sigma_i(\omega,t)|^{\rho_i}-\epsilon_i) &\text{if} \enskip \alpha_i > \alpha_{m,i}\\
\eta_i &\text{if} \enskip\alpha_i \leq \alpha_{m,i},
\end{cases}
\end{align}
where  $\bar{\omega_i}, \rho_i > 0$, $\epsilon_i$ and $\eta_i$ are positive constants and $\alpha_{m,i}$ is an adaptation threshold, chosen to be greater than $\alpha_{*,i}$.

In trying to find a condition for the convergence of the proposed control and adaptation schemes, let us consider the Lyapunov function candidate:

\begin{equation}\label{Eq19a0}
V = \dfrac{1}{2}\sigma^T I \sigma + \sum\limits_{i=1}^3\dfrac{1}{2\gamma_i}(\alpha_i -\alpha_{M,i})^2,
\end{equation}  
where $I$ is the inertia matrix, $\gamma_i$ is a positive constant, and $\alpha_{M,i}$ is the maximum value of the adaptive gain, i.e. $0<\alpha_{m,i}<\alpha<\alpha_{M,i}$. According to A.1, $\dot{I} = 0$. Thus, by taking the time derivative of $V$ and substituting $\dot{\sigma}$  from  (\ref{Eq15}), one has
\begin{align}\label{Eq19a1}\nonumber
\dot{V} =& \sigma^TI \dot{\sigma} +  \sum\limits_{i=1}^3\dfrac{1}{\gamma_i}(\alpha_i -\alpha_{M,i})\dot{\alpha}_i\\
=&\sigma^T\left( -I\ddot{X}_{1d}+ I\Lambda \dot{e} - S(\omega) I\omega + u + d \right) \nonumber\\
+&\sum\limits_{i=1}^3\dfrac{1}{\gamma_i}(\alpha_i -\alpha_{M,i})\dot{\alpha}_i. 
\end{align}
Equation (\ref{Eq19a1}) can be rewritten as,
\begin{align}\label{Eq19c}
\dot{V} &= \sigma^T (d + u_T)+\sum\limits_{i=1}^3\dfrac{1}{\gamma_i}(\alpha_i -\alpha_{M,i})\dot{\alpha}_i \nonumber\\
&= \sum\limits_{i=1}^3 \left[\sigma_i (d_i + u_{T,i}) + \dfrac{1}{\gamma_i}(\alpha_i -\alpha_{M,i})\dot{\alpha}_i\right]. 
\end{align}
For the case $\sigma_i\dot{\sigma}_i \leq 0$, from the twisting control law, we have
\begin{align}
\dot{V} &= \sum\limits_{i=1}^3 \sigma_i \left[d_i  -\alpha_i \mu_i\text{sign}(\sigma_i) \right] + \nonumber \\
&+ \sum\limits_{i=1}^3\dfrac{1}{\gamma_i}(\alpha_i -\alpha_{M,i}) \bar{\omega}_i \abs{\sigma_i}\text{sign}(|\sigma_i|^{\rho_i}-\epsilon_i) \nonumber \\
&= \sum\limits_{i=1}^3 \abs{\sigma_i}\mu_i \Big[\dfrac{d_i\text{sign}(\sigma_i)}{\mu_i} - \alpha_i \nonumber \\
&+\dfrac{\bar{\omega}_i}{\gamma_i\mu_i}(\alpha_i -\alpha_{M,i})  \text{sign}(|\sigma_i|^{\rho_i}-\epsilon_i)\Big].
\end{align}
By assuming that the disturbance $d$ 
is bounded, 
i.e., $|d_i| \le \Xi_{M,i}$, and with sufficiently small $\epsilon_i$ such that $|\sigma_i|^{\rho_i}>\epsilon_i$ \cite{Plestan:2010}, we have $\dot{V} \leq 0$ if
\begin{equation}\label{Eq19e}
\abs{\dfrac{d_i\text{sign}(\sigma_i)}{\mu_i}} \leq \alpha_i\text{ or } \alpha_i \geq \dfrac{\Xi_{M,i}}{\mu_i}.
\end{equation}
Noting that only the case $\alpha_i >\alpha_{m,i}$ is considered here as otherwise the last term in the right hand side of (29) becomes $\dfrac{1}{\gamma_i}(\alpha_i -\alpha_{M,i})\eta_i < 0$. For the case $\sigma_i\dot{\sigma}_i > 0$, from (23) we can have the same result as above if considering  $\mu_i=1$.

$ $

\section{Simulation and Validation} \label{results}
\begin{figure}[!htb]
    \centering 
    \includegraphics[scale = 0.42]{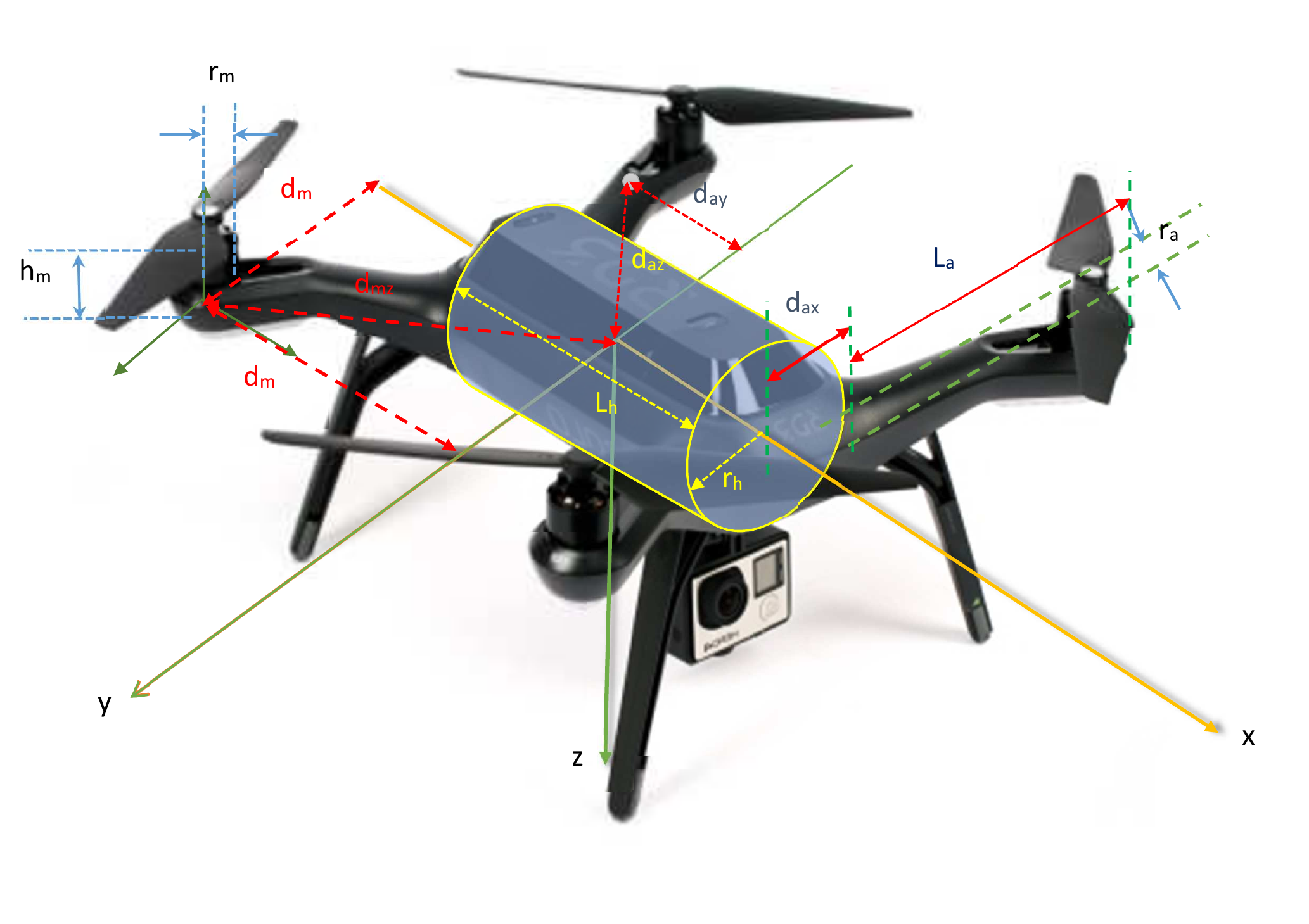}
    \caption{The 3DR Solo drone with body coordinate frame.}
\label{FigQuad}
\end{figure}

Extensive simulation and comparisons have been conducted to evaluate the performance of the proposed controller with the quadcopter model used for in this study being the 3DR Solo drone, shown in Fig. \ref{FigQuad}. It has three
processors, two are Cortex M4 168 MHz running Pixhawk firmware for low-level control and the other is an ARM Cortex A9 running Arducopter flight operating system. The UAV is equipped with a laser scanner, a camera and environmental sensors for data  acquisition. The programming is carried out and uploaded to the UAV through the ground control station called Mission Planner \cite{Hoang:2017}. The drone parameters obtained therein are listed in Table \ref{Table1}. The control parameters used for this study are given in Table \ref{Table2}.


\begin{center}

\captionof{table}{Parameters of the quadcopter model} \label{Table1}
	\begin{tabular}{ |c|c|c|c|c|c| }
		\hline
		Parameter 	& Value 				& Unit \\ \hline
		$m$ 	  	& 1.50  				& \textit{kg} \\
		$l$ 		& 0.205 				& \textit{m} \\
		$g$ 		& 9.81	 				& $m/s^2$ \\		
		$I_{xx}$ 	& $8.85\cdot 10^{-3}$  	& $kg.m^2$\\	
		$I_{yy}$ 	& $15.5 \cdot 10^{-3}$ 	& $kg.m^2$\\
		$I_{zz}$ 	& $23.09 \cdot 10^{-3}$ & $kg.m^2$\\
		\hline
	\end{tabular}
\end{center}

\begin{center}
	\captionof{table}{Control design parameters} \label{Table2} 
	\begin{tabular}{ |c|c|c|c| }
		\hline
		Variable 					& Value 		& Variable 									& Value \\ \hline
		$\lambda_1, \lambda_2$ 		& 4.68	 		& $\bar{\omega}_1, \bar{\omega}_2, \bar{\omega}_3$	& 200	\\
		$\lambda_3$ 				& 3.84			& $\alpha_{m,1},\alpha_{m,2},\alpha_{m,3}$	& 2.001		\\
		$\rho_1, \rho_2, \rho_3$	& 3.0			& $\alpha_{M,1},\alpha_{M,2},\alpha_{M,3}$	& 2.12		\\
		$\mu_1, \mu_2, \mu_3$ 		& 1/4	 		& $\Xi_{M,1},\Xi_{M,2},\Xi_{M,3}$ 			& 0.5 	\\
		$\epsilon_1, \epsilon_2, \epsilon_3$& 0.6	& $\eta_1, \eta_2,\eta_3$					& 0.01\\
		\hline 
	\end{tabular}
\end{center}
\vspace{12pt}
\subsection{Control performance in nominal conditions} \label{nomial}
\begin{figure}[!htbp]
\centering
\includegraphics[width=9cm]{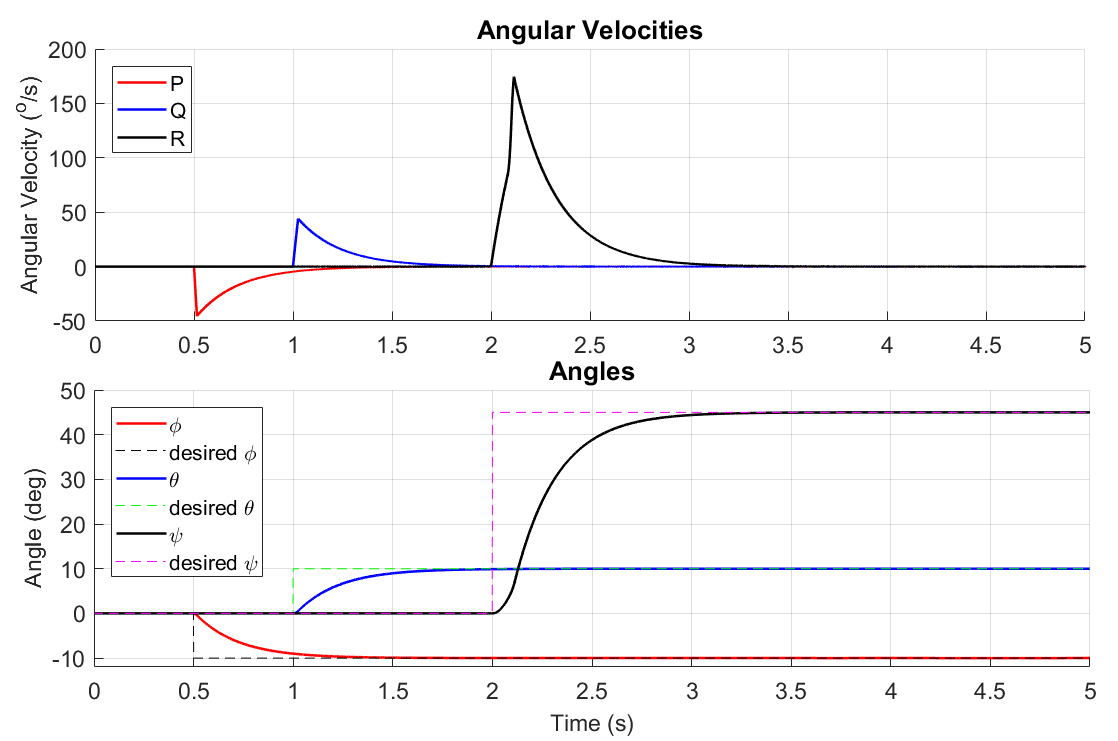}
\caption{Responses of the quadcopter ($P$, $Q$ and $R$- roll, pitch and yaw angular velocities) in nominal conditions.}
\label{FigNorminal}
\end{figure}

\begin{figure}[!htbp]
\centering
\includegraphics[width=9cm]{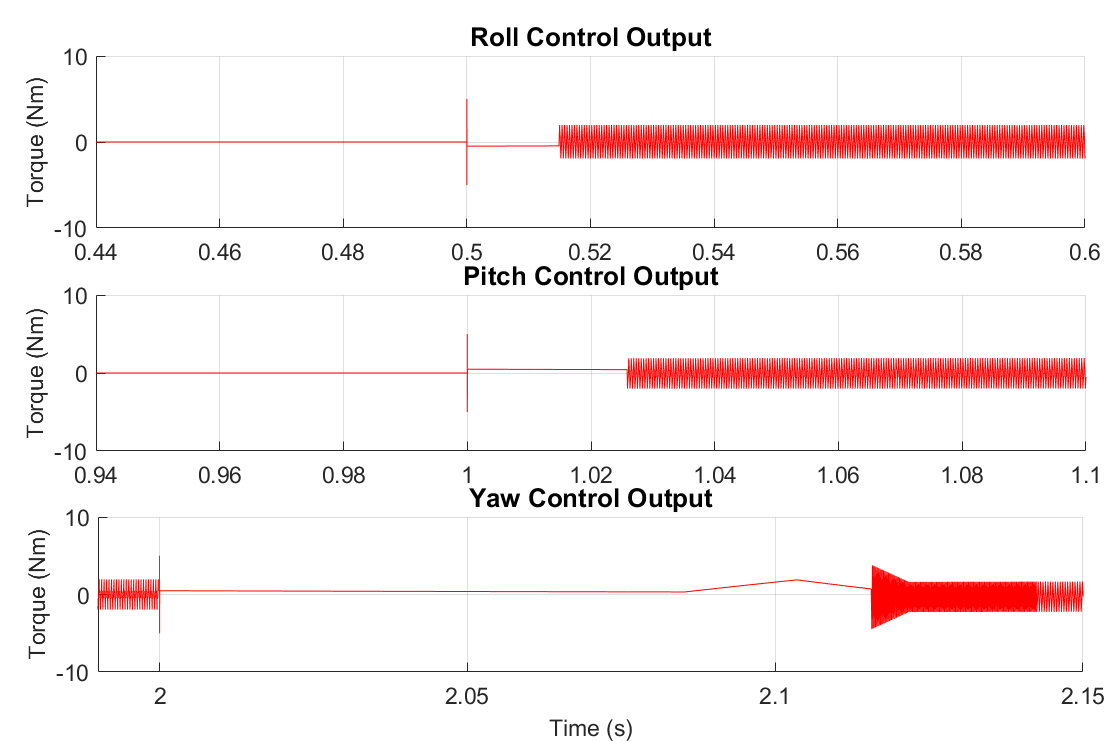}
\caption{Control torques.}
\label{Fig:torques}
\end{figure}

Performance of the controller is first evaluated in nominal conditions. In this case, the quadcopter is assumed to be at a hovering condition in a steady state where all attitude angles and angular velocities are zeros. New reference angles are then provided with the values $\phi = -10^\circ$, $\theta = 10^\circ$ and $\psi = 45^\circ$ at time 0.5 s, 1 s and 2 s, respectively. The system responses and controller outputs are shown in Fig. \ref{FigNorminal} and Fig. \ref{Fig:torques}, respectively, where the latter shows the zoomed-in time scale to observe the abrupt change in references and coupling effects. It can be seen that the proposed controller smoothly drives the angles to the reference values within one second and with a small overshoot despite strong coupling relations among control variables as described in (\ref{Eq6d}-\ref{Eq6f}). However, there still exists minor chattering from the numerical integration of the control system, as depicted in Fig. \ref{Fig:torques}. This can be interpreted as the trade-off to obtain a better control performance, which requires a larger gain $\alpha_i$ in (\ref{Eq17}). Nevertheless, this phenomenon can be mitigated by adaptively adjusting $\alpha_i$ to its threshold value in the steady state.

\subsection{Responses to disturbances} \label{disturbance}
\begin{figure}[!htbp]
\centering
\includegraphics[width=9cm]{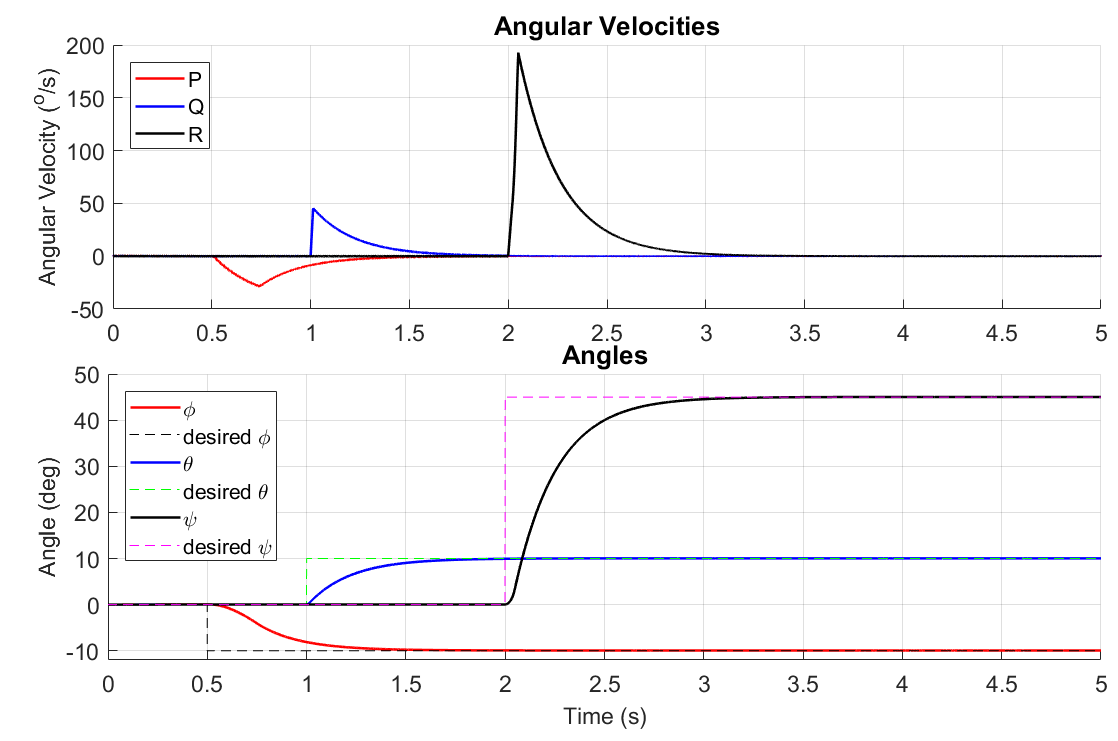}
\caption{Angular velocity and angle responses in the presence of disturbances.}
\label{Fig5}
\end{figure}

In this simulation, robustness of the controller is tested by adding disturbances with the mean value of 0.5 Nm to the torques in all three body axes of the quadcopter, corresponding to angular acceleration disturbances in (10-12). Reference values were selected to be the same as in the previous simulation. Results are shown in Fig. \ref{Fig5}. It can be seen that the proposed controller effectively rejects external disturbances to drive the quadcopter to reach the expected attitude within a similar time period as in nominal conditions.
\subsection{Responses to parametric variations} \label{variation}
\begin{figure}[!htbp]
\centering
\includegraphics[width=9cm]{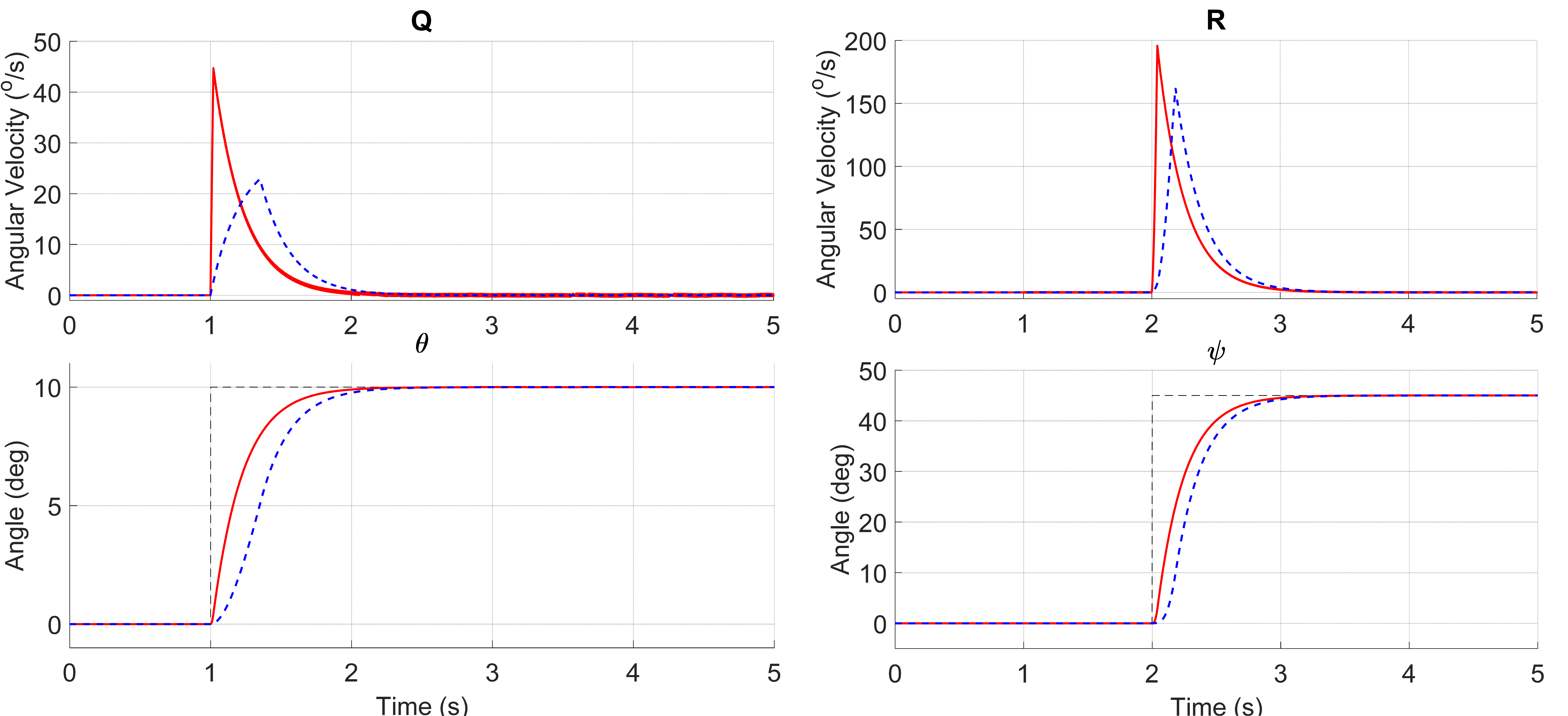}
\caption{Angle and angular velocity responses in the presence of parametric variations.}
\label{vary}
\end{figure}

In this simulation, the quadrotor is subject to several sources of uncertainties including variations in loads and moments of inertia. Specifically, a load of 0.8 kg, the largest load the 3DR Solo quadcopter can carry, is added to the model together with the following uncertainties in moments of inertia:
\begin{align}
\Delta I = \left[ \begin{array}{ccc}
0 & 0.0044 & -0.0077\\
0.0044 & 0 & 0.0115\\
-0.0077 & 0.0115 & 0\\
\end{array} \right].
\end{align}
Figure \ref{vary} shows the results in comparison with the nominal conditions. The settling time and overshoot of the responses are almost identical, indicating high robustness of the proposed controller. The variation of the adaptive gain observed in simulation is in the interval $2.001 \leq \alpha_i(t) \leq 2.12, i = 1,2,3$.  
Higher gain magnitudes imply more energy is required to stabilise the system to cope with the increase in disturbances and uncertainties owing to effectiveness of the adaptation.

\begin{figure}[!htbp]
\centering
\includegraphics[width=9cm]{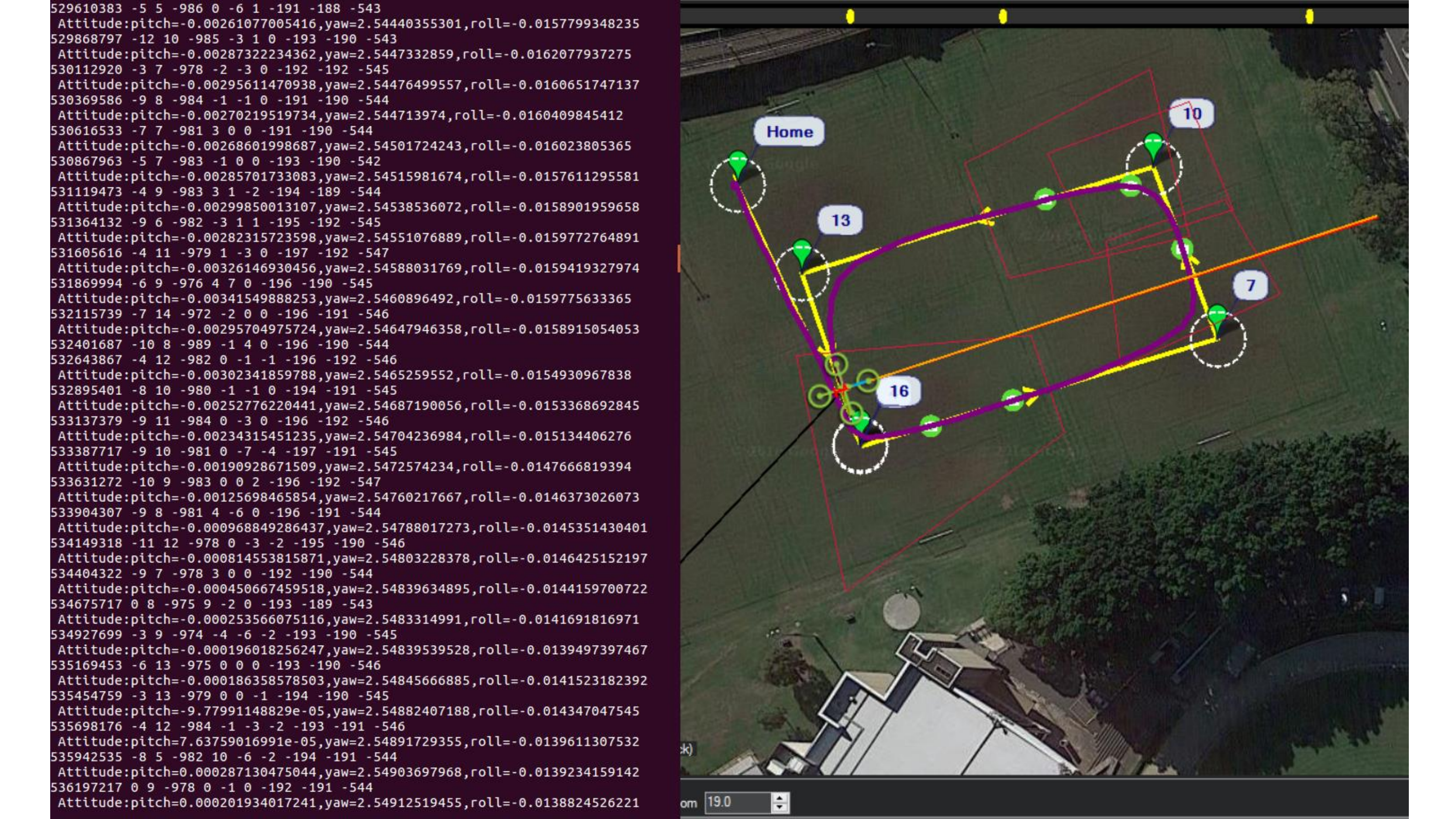}
\caption{Experimental data acquisition.}
\label{Experiment}
\end{figure}

\begin{figure}[!htbp]
	\centering
	\includegraphics[width=9cm]{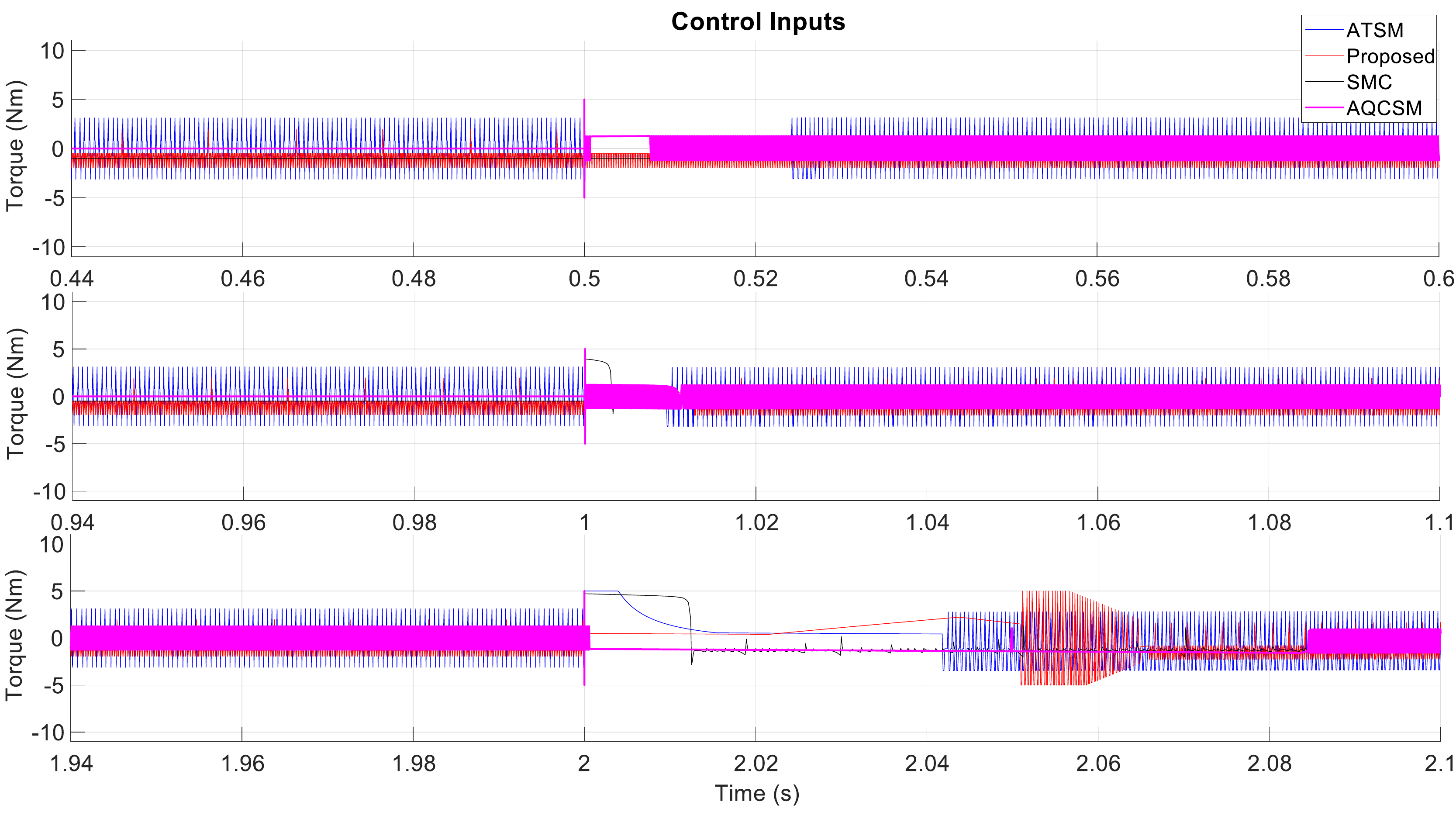}
	\caption{Time responses of three control inputs.}
\label{Diff_inputs}
\end{figure}

\subsection{Comparison and validation with real-time data} 


\begin{figure}[!htbp]
	\centering
	\begin{subfigure}[b]{0.56\textwidth}
		\includegraphics[width=9cm,height=4cm]{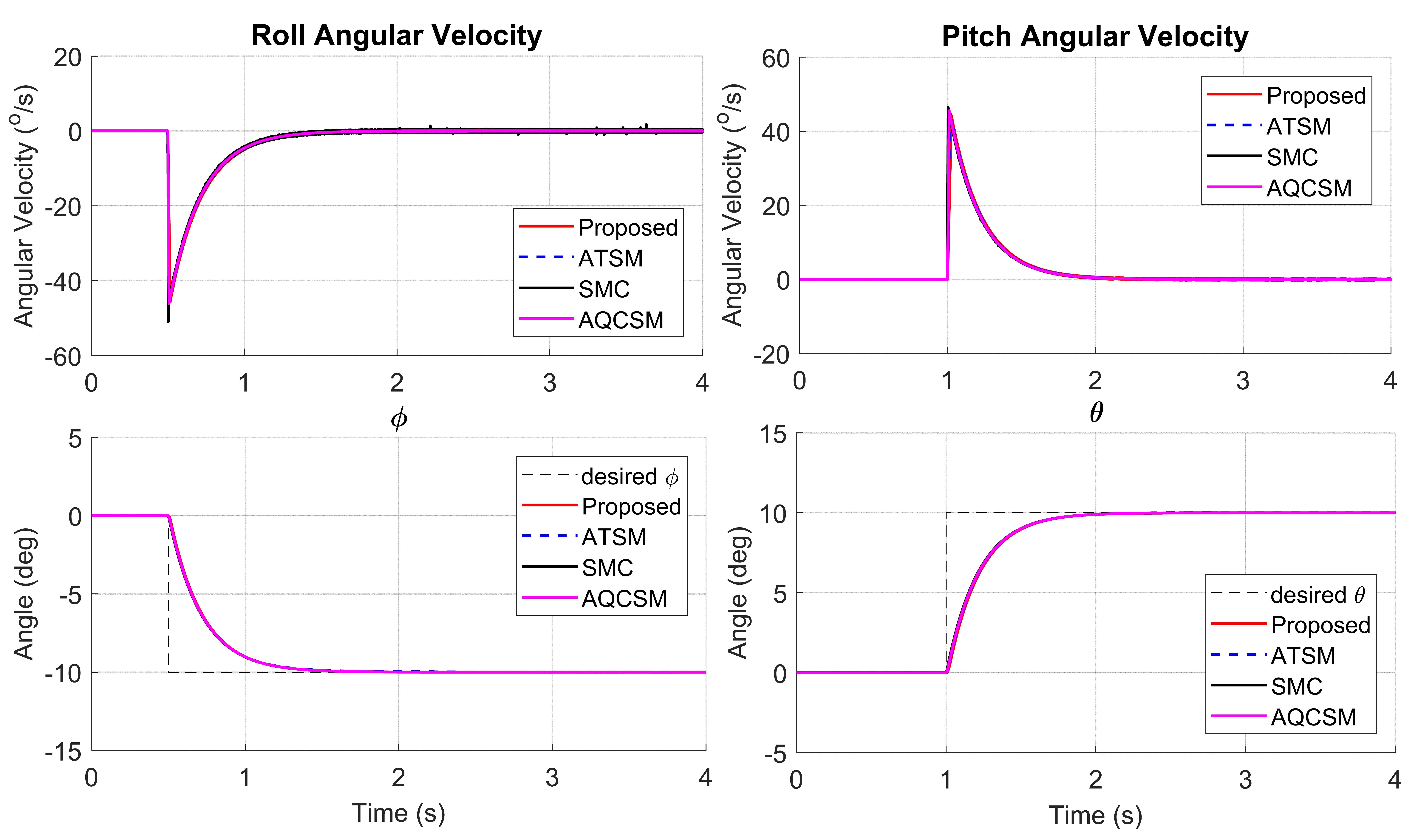}             
		\caption{}
		\label{state_comp:1}
	\end{subfigure}
	\centering
	\begin{subfigure}[b]{0.56\textwidth}
		\includegraphics[width=9cm,height=4cm]{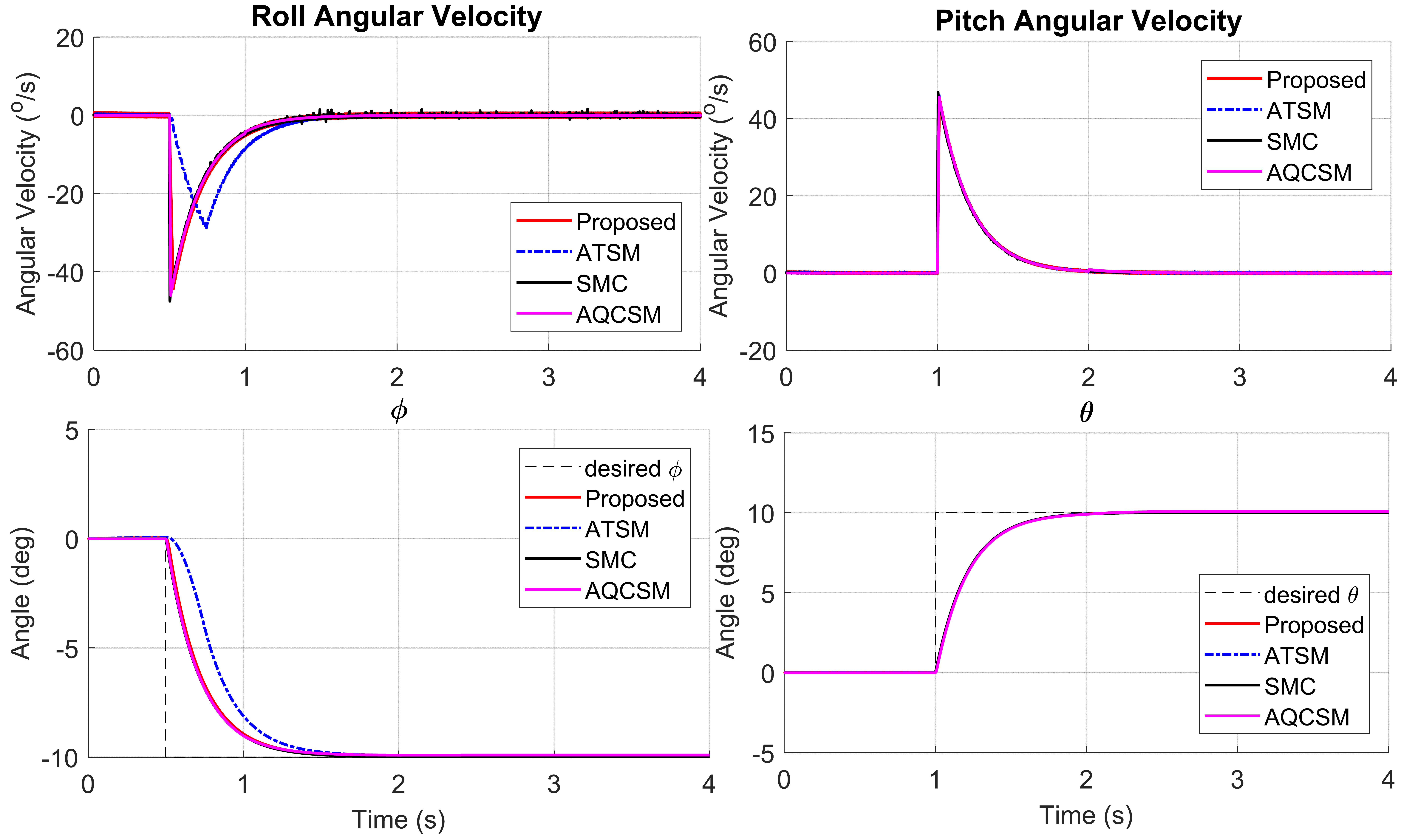}              
		\caption{}
		\label{state_comp:2}
	\end{subfigure}
	\centering
	\begin{subfigure}[b]{0.56\textwidth}
		\includegraphics[width=9cm,height=4cm]{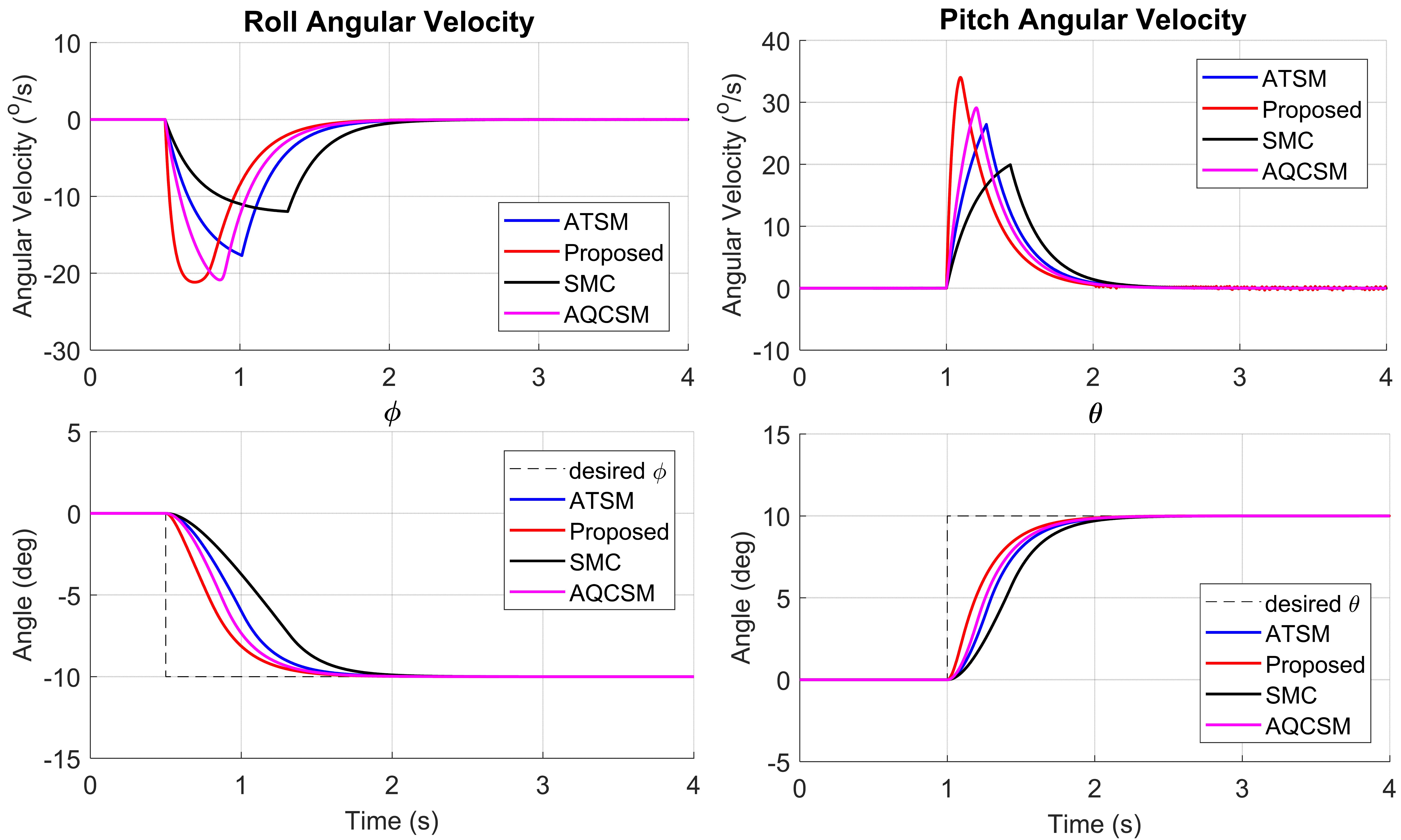}             
		\caption{}
		\label{state_comp:3}
	\end{subfigure}
   \caption{The roll and pitch angle and angular velocity responses of controllers in three scenarios:\\
 (a) Nominal condition; (b) Occurrence of disturbances; and (c) Parametric variations.}
   \label{state_comp}
\end{figure}
\begin{figure}[!htbp]
\centering
\includegraphics[width=9cm]{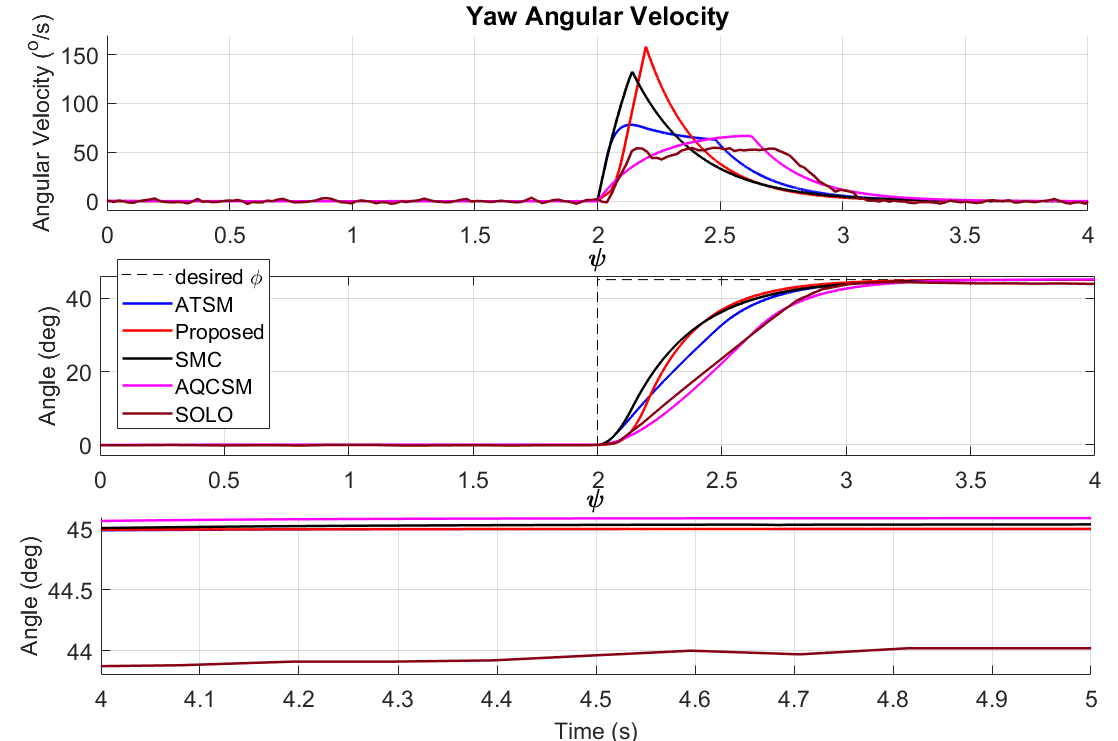}
\caption{Zoom-in tracking errors of controllers at the steady state.}
\label{YawZoomIn}
\end{figure}

For evaluation of the proposed control approach, simulation results were compared with real-time data obtained by using the built-in PID controller of the 3DR Solo drone to perform the mentioned attitude control. Figure \ref{Experiment} shows the flying path and data recorded, omitting position information, during the experiment. To compare performance of the proposed controller with other control techniques, the output responses were compared with those responses obtained by using the conventional SMC, the adaptive quasi-continuous (AQCSM) \cite{Hoang:2017} and the accelerated twisting sliding mode (ATSM) \cite{Dvir:2015}. The comparison was conducted under scenarios similar to the ones described in Section \ref{nomial}, \ref{disturbance} and \ref{variation}. Also, for the validation  purpose, simulation results were compared with real-time data obtained when the drone performing similar attitude control tasks. These results are shown in Fig. \ref{Diff_inputs} for the three control torques, where the proposed controller results in better tracking performance with reduced chattering. Indeed, Fig. \ref{state_comp} and \ref{YawZoomIn} show the time responses of Euler angles and angular velocities wherein the yaw tracking errors in the steady state are zoomed in. 
It can be seen that all controllers show similar performance as in nominal conditions. In the presence of disturbances, the proposed controller can however exhibit the smallest tracking errors, indicating a better capability of dealing with disturbances. In the case of parametric variations, the proposed controller can also provide the fastest convergence thanks to our proposed adaptive scheme.

\section{Conclusion}
In this paper, we have proposed an adaptive twisting sliding mode approach for robust control of quadcopter UAVs. The proposed controller is a modification of the accelerated twisting sliding mode control with an adaptive scheme to adjust the discontinuous gain to deal with not only external disturbances but also parametric variations. Performance of the controller is evaluated and compared with other controllers in various simulation scenarios. Its validity is also confirmed by comparing with experimental real-time data. Our future work will focus on implementing the proposed controller to enable higher level tasks of the drone such as cooperative tracking and visual inspection of infrastructure.



%

\bibliography{IEEEabrv,bibi}

\end{document}